\documentstyle[11pt,newpasp,twoside,epsf]{article}
\def\edcomment#1{\iffalse\marginpar{\raggedright\sl#1\/}\else\relax\fi}
\reversemarginpar

\begin{document}

\title{ Circumstellar Disks in pre-Main--Sequence Stars}
 \author{Antonella Natta}
\affil{Osservatorio di Arcetri, INAF, Largo Fermi 5, I-50125 Firenze, Italy}

\begin{abstract}
This  review  covers  the properties of disks around
pre-main--sequence stars. It is at this time in the evolution
that planets form, and it is important to understand the  properties of
these disks to understand planet formation.  I discuss
 disk shapes, masses and temperatures, the
properties of the host stars, disk lifetimes and dissipation processes.
Finally, evidence of grain growth during  pre-main-sequence evolution is
briefly summarized.
\end{abstract}

\section{Introduction}

Circumstellar disks surrounding pre-main--sequence (PMS) stars of solar 
and intermediate mass have been
known and studied, firstly indirectly, through their
unresolved emission in the near and far infrared and at millimeter 
wavelengths, and more recently by direct imaging in the optical, near 
infrared and millimeter. Excellent reviews of a number of topics concerning
disks in the pre-main--sequence phase of the evolution can be found in
Protostars and Planets IV (Mannings, Boss \& Russell eds.~ 2000).

The formation of circumstellar disks is considered a necessary consequence of
the process by which a molecular core collapses to form a star
(see Shu et al. 1987). However, disks live behind this first
period during which  stars  form.
PMS stars, which
have accreted most of their final mass and dispersed most of
their parental cloud, are still surrounded by
disks. We usually assume
 that such disks contain gas and dust with properties that do not differ much from those of the gas and dust in the parental core.

As the star contracts toward the main sequence, the disk ages. 
A variety of  processes  change the properties
of the gas and dust in the disk, and finally lead to  disk dissipation.
One  such process, but not the only one,
is the growth of  dust grains and the formation of planets. 
Planets form in disks around PMS stars,
and the properties of planetary systems
depend to a large degree on the properties of their birthsites,  the
PMS disks, and on how they change with time.

In this review, I will first summarize what we know of the 
properties of PMS disks around solar-type stars.
I will then address the question of what kind of stars have 
disks during their PMS phase, and how these disks  evolve. 
Finally, I will discuss the evidence for grain growth during the PMS disk
lifetime.

\section {Properties of disks around PMS stars}

Disks have been detected around PMS stars with
 masses between about 0.5 and 5 M$_\odot$
(T Tauri stars or TTS and Herbig Ae stars or HAe). Some of their
properties  are known from observation and  comparison with theoretical
models. However, there are still some
important parameters (for example, the run of the surface density
with radius) which
 are not well constrained by the existing observations. Also, as 
often, new and improved
 observations reveal unexpected and complex properties, that
need then to be taken into account.

\subsection {Shape}

An important recent result is the direct observation of disk shape
by means of
the HST observations of
disks driving HH jets (see McCughrean, Stapelfeldt \& Close 2000 and references therein;
the first image was obtained by Burrows et al. 1996 for the object HH30).  
HST images map the shape
of the disk surface seen in  scattered stellar light and show that
the disks  are ``flared", i.e., that the ratio $H/R$ of the  scale height 
over the distance from the star
is an increasing function of $R$. 
This is expected for gaseous disks in hydrostatic equilibrium 
(Kenyon \& Hartmann 1987), and in fact detailed analysis of the HST scattered light images has shown good quantitative
agreement between observations and theory. 
In a disk in hydrostatic equilibrium, the scale height $H$ at each radius is
determined by the equilibrium between the local pressure and the gravitational field of
the central star:
\begin{equation}
{{H}\over{R}} \>=\> \Big({{T_d}\over{T_c}}\Big)^{1/2} \> \Big({{R}\over{R_{star}}}\Big)^{1/2} 
\end{equation}
where $T_d$ is the disk temperature at $R$ (assumed isothermal in the vertical direction) and $T_c$ is a measure of the gravitational potential of the star:
\begin{equation}
T_c \> = \> \mu_g \> {{GM_\star}\over{k R_\star}} 
\end{equation}
where $\mu_g$ is the mean molecular weight of the disk material, $M_\star$
and $R_\star$ are the mass and radius of the central star, respectively. 
If the disk mass is dominated by gas, then $H/R$ is about 0.03 at 1 AU,
and 0.1--0.15 at 100 AU (for a typical TTS; Chiang \& Goldreich 1997).
On the contrary,
if the disk mass is dominated by dust, $\mu_g$ is much larger and $H/R$ 
becomes negligibly  small. Dust disks are expected to be geometrically
flat, rather than flared. 

During the PMS evolution, most of the disk heating is caused by irradiation from the
central star, and
the shape of the disk affects its temperature profile, and, therefore, its
spectral energy distribution (SED). Flared disks  intercept a larger fraction of 
the stellar radiation at large radii than flat disks, so that
their SED is a flatter function of wavelength (Kenyon \& Hartmann 1987;
Chiang \& Goldreich 1997). Most PMS stars have SEDs 
typical of flared disks (see, for example, Chiang et al.~2001; Natta et al.~2001), 
and we can reasonably conclude, even 
when direct images are not available, that they must have gas-rich disks.
Self-consistent calculations of the structure of
disks in hydrostatic equilibrium and their SEDs
have been computed, e.g., by D'Alessio et al. (1998, 1999, 2001)
Bell (1999), Dullemond, van Zadelhoff \& Natta (2002).

Note that both observations, namely the flared shapes seen in scattered light
and the flat SEDs, require not only gas-rich disks, but also that
relatively small (sub-micron)
grains are well mixed to the gas at all vertical heights,
i.e., that dust sedimentation and grain growth have not proceeded very far.
It is interesting to note that
there are some objects with SEDs typical of flat disks
(Meeus et al. 2001) or, at least, significantly less
flared  than predicted by hydrostatic equilibrium models (Chiang et al.~2001; 
D'Alessio et al.~2001). It is  possible that in these systems
grains have started settling toward the disk midplane,
separating from the gas. The gaseous disk is still flared, but the dust disk 
is much flatter. It seems unlikely, given that there is some indication of
gas accretion onto these stars, that they represent an even older stage
of the evolution, when gas has dissipated and dust dominates the disk mass.
These ``steep SED" objects deserve to be explored in more detail.

\subsection {Mass}

Another very important quantity for planet formation studies is
disk mass.
It is reasonably easy to measure the mass of the dust in the disk, from the 
flux at
some long wavelength (typically, in the millimeter) where the emission is
very likely optically thin. In its simplest formulation, the dust 
mass is  given by the expression:
\begin{equation}
M_{dust} \>= D^2 {{F_\nu}\over{\kappa_\nu B_\nu(T_{dust})}}
\end{equation}
where $\kappa_\nu$ is the dust opacity at the observed frequency, $T_{dust}$ the 
dust temperature (typically, 10-20 K in TTS), $D$ the distance and $F_\nu$ the observed flux.
This expression can be improved using more detailed disk models, which
 take into account
the temperature and surface density profile by fitting, for example, the spectral energy distribution over a wide range of wavelengths. 
For sources of known distance, the main uncertainty is due to the uncertainty on 
the dust opacity.
Most authors assume $\kappa_{1.3\rm{mm}}=0.5-1$ cm$^2$/g of dust, but
this value is very uncertain if significant grain growth has occurred
(see Beckwith, Henning \& Nakagawa 2000).

Once the dust mass has been determined, 
it is necessary to correct
for the gas-to-dust mass ratio. In disks, this is uncertain, and 
one assumes a  value of 100, which is typical of the interstellar medium.
Disk masses derived in this way range roughly from 0.003 to 0.3 M$_\odot$,
with a large spread 
even for stars of similar properties. Fig.~1 shows an updated
version of the compilation of Natta, Grinin \& Mannings (2000).

\begin{figure}
\plotone{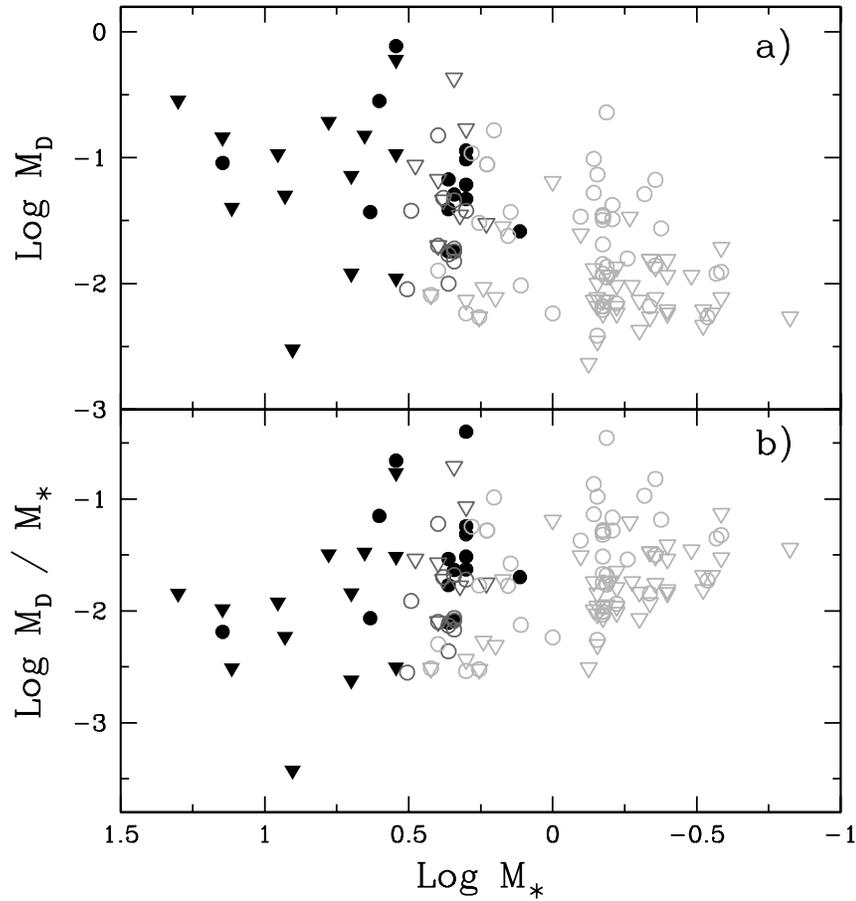}
\caption{Top panel: disk masses (gas+dust, assuming a gas-to-dust mass ratio of 100)
derived from millimeter fluxes are plotted as 
function of the stellar mass.  Triangles are upper limits, dots detections;
filled symbols are interferometer data, open symbols are single-dish
measurements. The lower panel plots the ratio of the disk to the stellar mass.
Data from the compilation of  Natta et al.~(2000), with some additional points
for early B stars from OVRO (Mannings et al.~2000, personal communication) and
IRAM/PdB (Fuente et al.~2001).
}
\end{figure}

Direct determinations of the gas mass exist for some objects, but they do 
suffer from large uncertainties.
A number of  PMS disks have been observed in gas  tracers such as CO.
Gas masses derived from CO, using standard values of
the conversion factor CO/H$_2$,
are  much lower (by orders of magnitude) than those obtained from dust
(Guilloteau \& Dutrey 1998; Mannings \& Sargent, 1998, 2000).
It is likely that the molecular tracers are affected by depletion onto grains
(see, for example, van Zadelhoff et al.~2001), and,
in some cases, by optical depth effects. 
Observations of several molecular species and transitions can be used to
constrain the local gas density and to infer from that the disk mass.
This has been done so far for only two objects (Dutrey, Guilloteau \& Guelin 1997), and
showing that depletion occurs in several molecules.
Recently, some
observations of mid-infared H$_2$ lines have been obtained with ISO
for a handful of TTS and HAe stars, and the results seem to indicate a better
agreement with dust-derived disk masses (Thi et al. 2001);
however, the observed H$_2$ lines do not
trace the cold gas, which is likely to dominate the mass, so that large extrapolations 
are required. Also, in spite of the careful selection of
objects aimed at excluding those with surrounding low-density gas,
 there is some doubt that the H$_2$ emission has its origin in the disk,
since ground-based observations obtained with a beam much smaller than
that of ISO do not detect any comparable emission 
(Richter \& Lacy, this conference).
Future space missions, like SIRTF, will certainly   clarify this issue.

\subsection {Size}

Disk sizes can be directly obtained from images at various
wavelengths. As one might expect,
the result depends on the tracer and on the wavelength of the
observations. Millimeter interferometers (PdB, OVRO, VLA at 7mm) 
have been used to map the outer disk both in the dust continuum emission
and in molecular gas, mostly CO.
Dust millimeter emission is
generally rather compact; when resolved, disks have radii 
(HWHM) of 50--100 AU  ( see Wilner \& Lay 2000 and references therein
and Mannings \& Sargent 1998, 2000).
Maps in the lines of various CO isotopes  show
larger radii, typically
a few hundred AU (Koerner \& Sargent 1995;
Dutrey et al.~1996; Mannings \& Sargent 1998, 2000). The HST images in scattered light give similar 
sizes, but some
disks in Orion seen as silhouettes against the luminous background of
the Orion nebula can be as large as 1000 AU (McCaughrean \& O'Dell 1996).
It is likely that these apparent differences result from a combination
of the dependence of the specific tracer on the distance from the star 
with the sensitivity of the  adopted technique.

To summarize, we can safely say that there is
clear evidence of gas in keplerian rotation on scales
of few hundred AU (see Dutrey 2000). In some objects,  matter in
a flattened structure is present  even
at larger distances from the central star.

\subsection {Temperatures}

\begin{figure}
\plotone{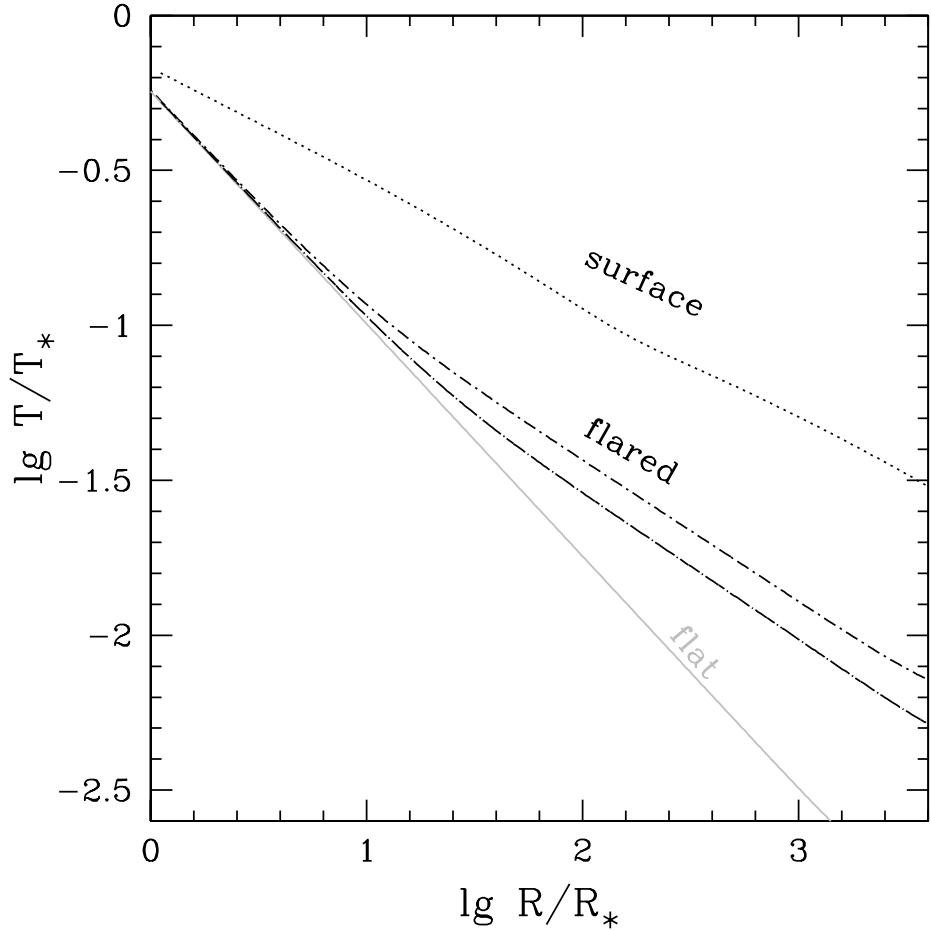}
\caption {Ratio of the disk temperature to the
temperature of the central star  as function of $R/R_\star$. 
The two dot-dashed curves are for flared
disks, one for a disk around a Herbig Ae star (dot-long dashed), the other for a disk around
a M7 star (dot-short dashed). The figure shows also the temperature profile of a flat disk
(solid  line) and of the disk surface (dotted line); 
these two quantities, normalized to $T_\star$, are practically independent
of the central star.
} 
\end{figure}

The disk thermal structure  can only be derived by comparing observations
with model predictions. In principle, we would like
to know both dust and gas temperatures, which are coupled only in the high-density
regions of the disk, and to determine for both  the variation of $T$ in the radial
and vertical directions. 

At the moment, the most abundant information is for the run of dust temperature
with radius, which is obtained by comparing model predictions to measured SEDs.
Most PMS disks are heated by irradiation from
the central star, and dissipation of 
viscous energy may play a role, if any, only in the very inner disk
(Calvet, Hartmann \& Strom 2000). Therefore, to a good approximation
the disk temperature $T$
is proportional to the stellar temperature, ranging from about 1/2 $T_\star$ near the
star to about 0.01 $T_\star$ at $10^3$ $R_\star$ (for flared disks), as shown in Fig.~2.
This temperature is in fact an effective temperature, computed by balancing the
absorbed and emitted energy at each radius. To zero order,
the disk is isothermal in the vertical direction, so that
it is generally assumed that  $T$ is the temperature of the
disk midplane, where most matter resides.

However, simple considerations show that in an irradiated disk the surface
is hotter than  the midplane (Calvet et al.~1992, Chiang \& Goldreich 1997), so that
at each radius there is a vertical gradient of temperature, which 
does indeed  affect the shape of the SED and produce dust emission features,
as observed in most TTS and HAe stars
(see, for example, Chiang \& Goldreich 1997). 

The gas temperature is likely to differ from the dust temperature as soon as the density
drops below some critical value (typically, of order $10^4-10^5$ cm$^{-3}$). This
is likely the case in the surface layers of the disk, as well as in the disk midplane
at large distances from the star.
Since molecular excitation and 
chemistry in the disk depend on  the gas temperature,
predictions of the expected molecular line
intensities depend on a good knowledge
of the disk vertical structure and of the gas-dust thermal coupling.
This is an area of fast progress, both theoretically
and observationally, from which we can expect a wealth of results in the 
next few years.

One final point I would like to mention is
that the thermal history of the disk
material during the PMS life of the star can be very far from smooth,
as high-energy flares, activity-related  phenomena, shocks from
residual accreting gas and winds take place.
These aspects still need  exploratioationn.

\section {The disks of Herbig Ae stars}

One of the results of direct disk imaging is that
PMS disks are 
not smooth, simple structures. They are perturbed by a large variety of effects
(for example, dynamical perturbations by companions), in ways that can 
affect their evolution.
Sometime, high-spatial resolution observations seem to conflict with our
previous understanding of disks. Often, however, they may give us clues to long-standing
puzzles, as in the case of disks around Herbig Ae stars.

All the isolated Herbig Ae stars, i.e.,
PMS stars of mass 2--3 M$_\odot$ with little evidence of surrounding
matter, have a strong IR excess, reminiscent 
in many ways of that observed in TTS. However, the origin of this emission 
has been at the center of a controversy, which began with the 
work of Hillenbrand et al. (1992)
and lasted for many years (see Natta et al.~2000 for a summary).
At the core of the discussion was the fact that 
these stars emit in the near-IR a significant fraction of their luminosity (up to 25\%;
see Meeus et al.~2001), with   a peak at about 2--3 $\mu$m (the near-IR ``bump";
Hillenbrand et al.~1992). These  two
properties (shape of the near-IR bump
and energetics) are inconsistent with any standard disk model
(either heated by viscosity or by irradiation from the star), and there have been
various suggestions in the literature that in fact the IR emission of HAe stars
is not due to matter in a disk but rather to a more spherical shell of dust
(see, for example, Berrilli et al.~1992), possibly
containing very small, transiently heated grains (Hartmann, Kenyon \& Calvet 1993
).
Further doubts on the disk hypothesis were shed by
the first near-IR interferometric observations  published for some HAe stars
(Millan-Gabet, Schloerb \& Traub 2001), which show that the data are not consistent with the
predictions of standard disk models, but suggest instead that the emission comes from
a ring of dust at a distance from the star comparable to the dust sublimation 
radius (about 0.3 AU for the A0 star AB~Aur).

\begin{figure}
\plottwo{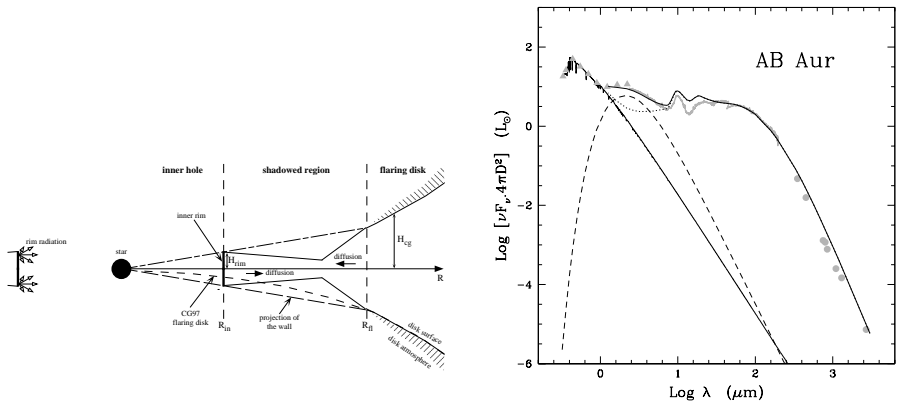}{ps.abaur_gap}
\caption {The sketch on the left (from Dullemond et al.~2001) shows the proposed
structure
of Herbig Ae disks. Note the inner hole, where only optically thin matter exists,
the rim, at the dust condensation radius, the shadowed region and the flared outer disk.
The right panel shows the SED of the star AB~Aur (from Natta et al.~2001). 
Triangles, dots and the thick line
are the observed data. The thin solid line is the stellar photosphere, the dotted line
is the SED of a flared, irradiated disk with no inner rim, and the dashed line
shows the approximate contribution of the rim. The total fits the observation rather well.
}
\end{figure}

Recently, it has been proposed that the IR properties of these HAe stars can be
understood if one considers the modification to the inner
disk structure introduced by dust sublimation. In a disk where the gas  is optically thin
to the stellar radiation, as is likely to be the case
in most optically visible HAe stars where the
accretion rate is very low (Grady, this conference), 
the condensation of dust at radii of few tenths of AU
from the star forms an optically thick inner rim which is frontally heated by the
star and is therefore hotter than the 
disk midplane.  The disk ``puffs up" and forms an inner rim
(Natta et al.~2001), whose
emission has the characteristics of the observed near-IR
excess (see Fig.~3). 
The SED of puffed up disks has been computed self-consistently (Dullemond,
Domink \& Natta 2001) and 
fits very well the observations. Also, it is possible to explain with these
models, the interferometry results, since the near infrared emission 
comes entirely from the inner rim.

A similar ``puffing up"  may occur also if there is a gap opened in the disk, for 
example by the formation of a planet (Sasselov \& Lecar 2000). In this case
also,  the dust at the
edge of the gap further from the star is  exposed frontally to the stellar radiation and
is therefore hotter than dust in the disk midplane at the same distance from the star.
In a recent paper Calvet et al. (2002) model the disk of the nearby
TTS TW~Hya as having an inner  gap and a puffed-up rim at  about 3--4 AU from the star,
possibly caused by a still growing protoplanet. As for the inner disks of 
HAe stars, interferometry in the near and mid-IR with the instruments soon coming on line,
such as VLTI, will provide direct images of such structures.

\section{Do stars of all masses have disks?}

As I have already mentioned, most of our knowledge of PMS disks is restricted to
 stars in a range of mass  roughly 0.5-5 M$_\odot$.
It is worthwhile to explore  disks around stars outside 
this range, much more and much less massive,
and for a variety of reasons. One, and the most obvious, is that
disks are the signpost of
star formation  via  gravitational contraction of
molecular cores; their presence can shed light on the stellar formation
mechanism. 
There is, however,  an additional reason to study disks around  ``extreme" objects,
namely that the very different physical conditions expected  in these disks
can help  clarify the various processes that  control disk evolution,  
including   planet formation.

\subsection {Herbig Be stars}
The process of formation of massive stars has been  the center of a long-lasting
debate (see Stahler, Palla \& Ho 2000 for a review). As an alternative to gravitational
collapse, where radiation pressure prevents the formation of stars above a certain
mass, it has been proposed that massive stars form by coagulation of lower mass
objects (Bonnell, Bate \& Zinnecker 1998). One prediction of this mechanism
is the lack of massive disks around massive stars.

It is very difficult to collect a significant sample of O stars where disks can be studied.
The best one can do is to look at
the most massive among Herbig Be (HBe) stars.
They have masses of about 8--20 M$_\odot$ and are very young
(less than few million years). These stars lack a PMS phase and when they become
visible, they are already on the ZAMS (Palla \& Stahler 1993).
About a dozen have been observed with millimeter interferometers and only one (R Mon)
has been detected. Fig.~1 shows that the derived upper limits to the ratio of the
disk to stellar mass (assuming the usual factor 100 to convert from the observed dust mass to total mass)
are significantly lower than the values of lower-mass stars.
Early B stars reach the
ZAMS with disks of relatively little mass, if any.

The conclusion that HBe stars  never  had disks is, however, premature. 
All evidence, including the presence of outflows and jets
and tentative detection of disks in the
embedded phase (see Kurtz et al.~2000),
 the distribution of the circumstellar gas in HBe stars of
different age
(Fuente et al. 2001, 2002),
and
the presence of extended dust structures around these objects
(Natta et al.~1993, Di Francesco et al.~1998, Abraham et al.~2000) 
points to a fast evolution  of
 the circumstellar matter. It seems likely that the HBe disks are quickly 
dispersed.
I will come back to the HBe case in  the following section.

\subsection {Brown dwarfs}

Recently, at the other extreme of the mass
range, the formation process
of brown dwarfs (BD) has become of interest, as more and more BDs are discovered
in star forming regions. Formation
theories alternative to  collapse of a molecular core
include the possibility that below a given mass, BDs form like giant planets
in gravitationally unstable regions of circumstellar disks
(Papaloizou \& Terquem 2001),
or that they are stellar embryos ejected from multiple systems (Reipurth \& Clarke 2001; 
Bate, Bonnell \& Bromm 2002).
In both these alternative views, one expects only very small, short-lived disks, 
if any.
The search for BD disks is producing some interesting results.
Using near-infrared color-color diagrams,  as for TTS,
Muench et al. (2001) find that 63\% of candidate BDs
in the Trapezium cluster
have disks.
However, in the older region $\sigma$ Ori (age 2--7 Myr), the same technique shows
that no BD candidate has disks (Oliveira et al. 2002).
Mid-infrared ISOCAM surveys have detected many candidate BDs associated with
 warm dust
in star-forming regions (Persi et al.~2000, Bontemps et al.~2001). As for the Trapezium objects,
the BD nature of the candidates needs to be confirmed spectroscopically.
More detailed studies of individual objects (Natta \& Testi 2001; Testi et al.~2002;
Natta et al.~2002) selected among mid-IR sources
in the young regions Cha I and $\rho$ Oph have shown that there are bona-fide BD with
excess mid-IR emission that can be modeled as being due to disks irradiated by the central object
(see Fig.~4). These disks are optically thick at mid-infrared wavelengths, and in some cases
(the two with the lowest mass in the sample), there is evidence that the disks have to be 
flared, i.e., gas-rich. The
SIRTF/Legacy observations of star forming regions, with
good wavelength coverage and sensitivity,
should allow us to explore the properties of BD disks in  detail.

\begin{figure}
\plottwo{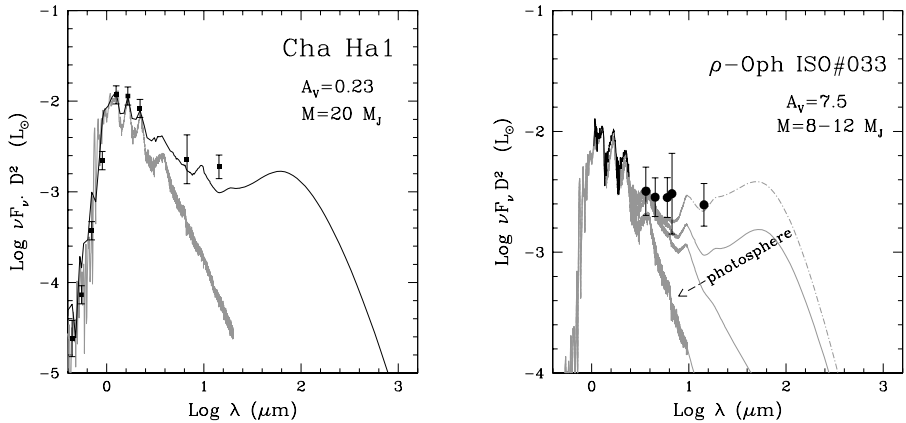} {ps.033}
\caption {Spectral Energy Distribution of two BDs, one in Chamaelion I (left panel,
mass about 0.02 M$_\odot$ or 20 Jupiter masses; from Natta \& Testi 2001)
and one in $\rho$ Oph (right panel, mass about 8-12 Jupiter masses; from Testi et al.~2002).
Dots and the thick black line for \#033 are observations. Thin grey lines show the
predictions of irradiated disk models of various flavours: solid lines are for irradiated
flared and flat disks extending to the stellar radius, dot-dashed line for a flared disk 
with a inner hole of about 3$R_\star$. For Cha H$\alpha$1, only the flared disk
with no inner hole is shown. To all models we have added the contribution of the
photosphere.}
\end{figure}

\section {Disk lifetime}

The lifetime of PMS disks is a crucial parameter, since any planetary system must
be formed before the  disk dissipates.
The current wisdom is that PMS disks live a few million years.
This is based on near-IR surveys of TTS in a number of young clusters of different age
(see Alves et al.~2002), and, strictly speaking, refers only to the inner disk,
which emits at those wavelengths. However, there is some  evidence (which, however,
should be re-discussed with the better statistics available today) that the process
of disk dissipation, once started, occurs rapidly
 (few $10^5$ yr;  Skrutskie et al.~1990), 
and over a similar timescale at all
radii (see, for example, Duvert et al.~2000),
in both dust and gas. 

There is  evidence that disks around more massive stars have shorter lifetimes. 
We have already mentioned the case of HBe stars, but the trend of rapid
 evolution seems
to extend to stars of lower mass (A and F), when statistical indicators are used. 
For example, in the Trapezium cluster Lada et al. (2000)  find that
only 42\% of stars of spectral type OBA have a near-IR excess, versus an average 78\%;
in IC 348 Haish, Lada \& Lada (2001) find that  0\% of stars of spectral types BAF have excess, versus 54\% of the total.

Although on average there is a clear trend of
disk frequency with age, individual objects tell us a much more complex story. 
There are (relatively) old stars (age $\sim$ 10 Myr) that still retain their PMS
gas-rich disk
(see, for HAe stars, Natta et al.~2000),
and young objects that have already lost it, as is the case for
many  weak-line T Tauri stars (WTTS).

This is in fact not surprising, when we consider the variety of processes that can
cause the dispersal of disks.  Among them are processes intrinsic to the disk itself,
such as viscosity and planet formation, and others that depend on the  interaction of the
disk with the central star, in particular with the radiation field and wind.
The environment where the disk evolves is also important. For example, 
a nearby massive star may evaporate the disk,  as in Orion,
or destroy it by tidal interactions. Stellar encounters may destroy the disk,
and
dynamical interaction between the disk and possible companions may also dissipate
it. 
It is also possible that stars are born with disks of different mass.
If, for example, 
the measured spread of disk masses seen in Fig.~1  is intrinsic, 
we should expect a similar  spread in disk lifetimes (more than a factor 100). It is
not impossible that WTTS
are objects born with disks at the lowest end of the mass spectrum, or
that the only disks which live long enough to form planets
are those at the upper end of the disk mass distribution. We do not know enough,
at present, to have a clear understanding of these points, but they are certainly worth
investigation  in the future.

Disk dispersal 
processes have been described in detail by Hollenbach, Yorke \& Johnstone (2000), and the
effect on disks of dynamical processes  is further discussed
by Artimowicz in this conference.  I will  confine myself here
to the case of Herbig Be star disks.

\subsection {The case of Herbig Be disks}

All disks evolve as matter is accreted inward and
angular momentum is transported outward. This process is usually described
in terms of viscosity, by which most of the mass accretes onto the star, while a very small fraction moves to infinity carrying the
disk angular momentum. 
One of the characteristics of viscous evolution is that the disk radius increases with time,
so that the viscous timescale of the outer regions, where at any given instant
the mass resides, keeps
increasing, and viscosity alone cannot explain the complete disk dissipation.

The disk evolution is also affected by the
presence of the central star, which
can cause photoevaporation of its surrounding disk. 
Evaporation by
photoionization  is irrelevant for TTS, and not very effective in HAe stars. In these
objects, however, softer photons ($12< h\nu <13.6$ eV) can photodissociate and warm up
the disk material and dissipate it in a  manner similar to photoionizing photons. 
Also, flares and the
UV radiation emitted in TTS by accretion shocks can erode the outer disk.
Hollenbach et al. (2000) conclude
that  photodissociation  may be able to dissipate the outer disks
of TTS  on timescales comparable to those observed. 

The case of HBe stars is particularly interesting, since these stars  have the
shortest timescale for disk dissipation of all PMS stars (much less than 1 Myr).
Even if one cannot exclude that other processes 
such as disk ablation by a stellar wind,
may have an important role, it is possible to explain this 
evolution simply by  the combined action of  two processes,
viscosity and photodissociation. Fig.~5 plots the viscous dissipation and
photodissociation  timescales for  a B0 ZAMS disk. One can see that the outer
disk (for $R$ greater than about 1/3$R_{cr}\sim 30$ AU, where
$R_{cr}$ is the  radius at which the gravitation
of the star balances the sound speed of the ionized disk material)
evaporates very rapidly ( in about $10^5$ yr). The residual disk may then be destroyed
by viscosity on a comparable timescale.


A combination of various processes is likely to occur in all stars, and may in
fact determine the characteristics of disk dissipation.
For example, Clarke, Gendrin \& Sotomayor (2001)
combine viscous evolution and photoevaporation by X-ray emission near the photosphere
to explain not only the typical lifetimes but also the rapid transition of TTS
from disk to diskless.

\begin{figure}
\plotone{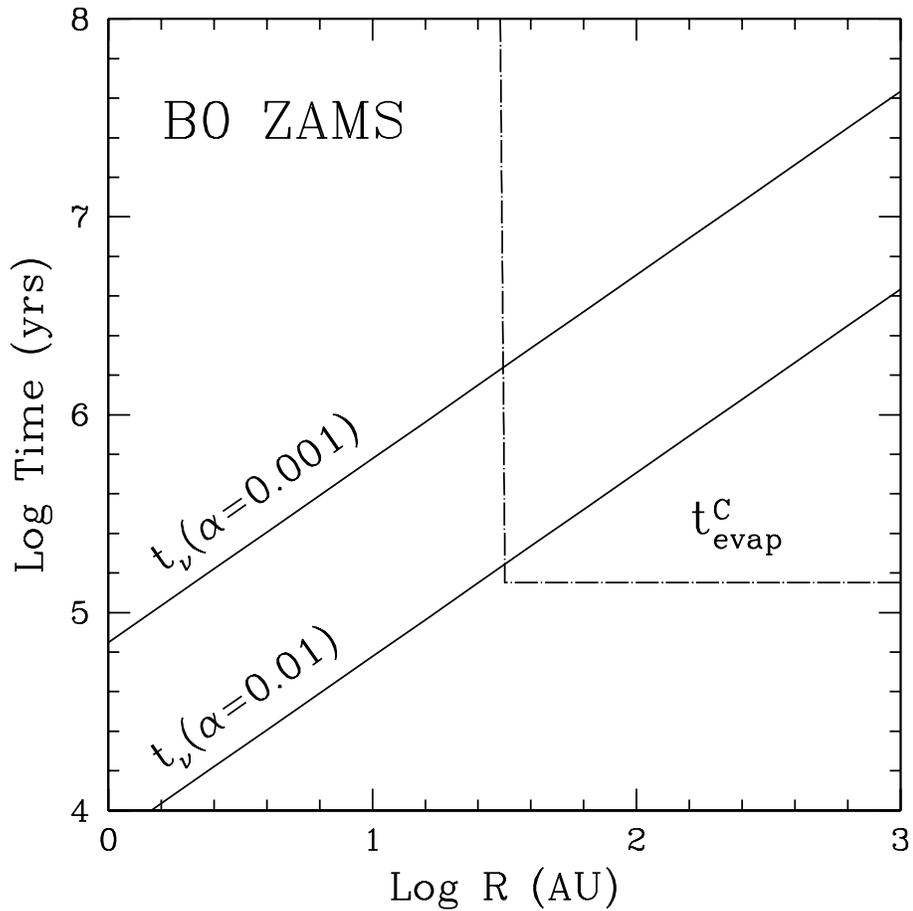}
\caption{Dissipation timescale for the disk of a B0 star as
function of the disk radius. The lines refer to the  photoevaporation caused
by the radiation field of the central star (dot-dashed line) and viscosity
(solid lines)
for two different values of the viscosity parameter $\alpha$,
as labelled. The disk mass is 0.5 M$_\odot$,
about 0.03 $M_\star$.}
\end{figure}

%



\section {Grain growth in PMS disks}

The last topic I want to discuss is that of grain growth in PMS disks.
Growth from the typical size of grains in the ISM 
(roughly smaller than few tenths of microns) to much larger values is expected to occurr in the
denser layers of circumstellar disks during the PMS phase.
This, if proved, would be a strong indication that the sequence of processes
we think lead from typical PMS disks to planets is indeed occurring.
However, a robust indication that PMS disk grains are growing has been
elusive.

The argument in favour of large grains in the disk midplane is an old one, and is based on a very simple argument. Suppose that the long wavelength
emission of a disk is optically thin, then the flux dependence on wavelength
should mirror the opacity law. Namely, if $\alpha$ is the observed spectral index 
($F_\nu \propto \nu^\alpha$) and the opacity can be expressed
as $\kappa_\nu \propto \nu^\beta$, then one should have
$\alpha = 2+\beta$. As soon as the first submillimeter and millimeter
observations became available, it was realized that TTS have lower
values of $\alpha$ than more embedded young stellar objects (see Fig.~6).
This result was interpreted as evidence for large grains in the TTS disk
midplane,
of sizes $>$ few millimeters. Note that this is not
in contradiction with the presence of silicate emission features
in the same objects, since such features are due to grains on the
disk surface, and a grain size vertical stratification is
expected. This argument, and its caveats, has been discussed, among others, by
Beckwith et al. (2000).

\begin{figure}
\plotone{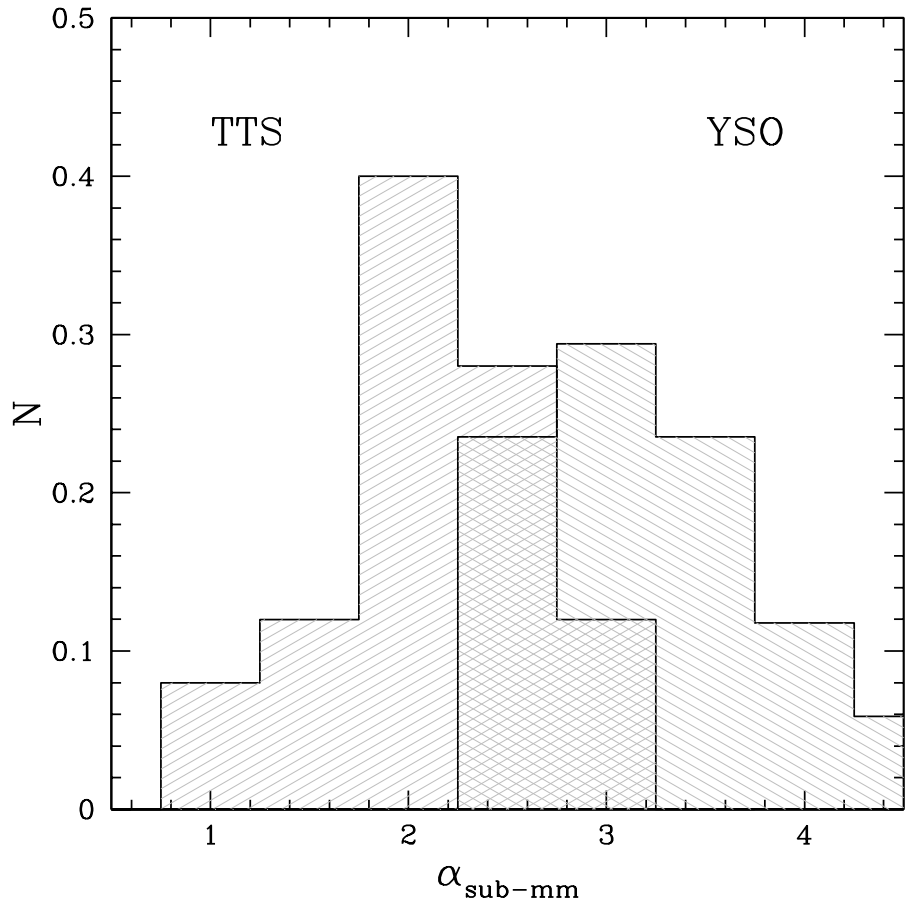}
\caption {Spectral index from sub-millimeter and millimeter photometry
for a sample of TTS (slashed hatching)
and more embedded YSOs (back-slashed hatching); data from Beckwith \& Sargent (1991)
and Moriarty-Schieven et al. (1994).}
\end{figure}

The main weakness of this result lies in the assumption that the disk
emission is optically thin. Indeed, if this is not true and the
disk remains optically thick over the whole observed wavelength range,
its emission would be that of black-body ($\alpha =2$), mimicking
the characteristics of extremely large grains with $\beta$=0.
This effect has been investigated in detail for few stars,
using interferometric data between 1.3mm and 7mm (Testi et al. 2001).
The results are shown in Fig.~7 for four stars. 

\begin{figure}
\plotone{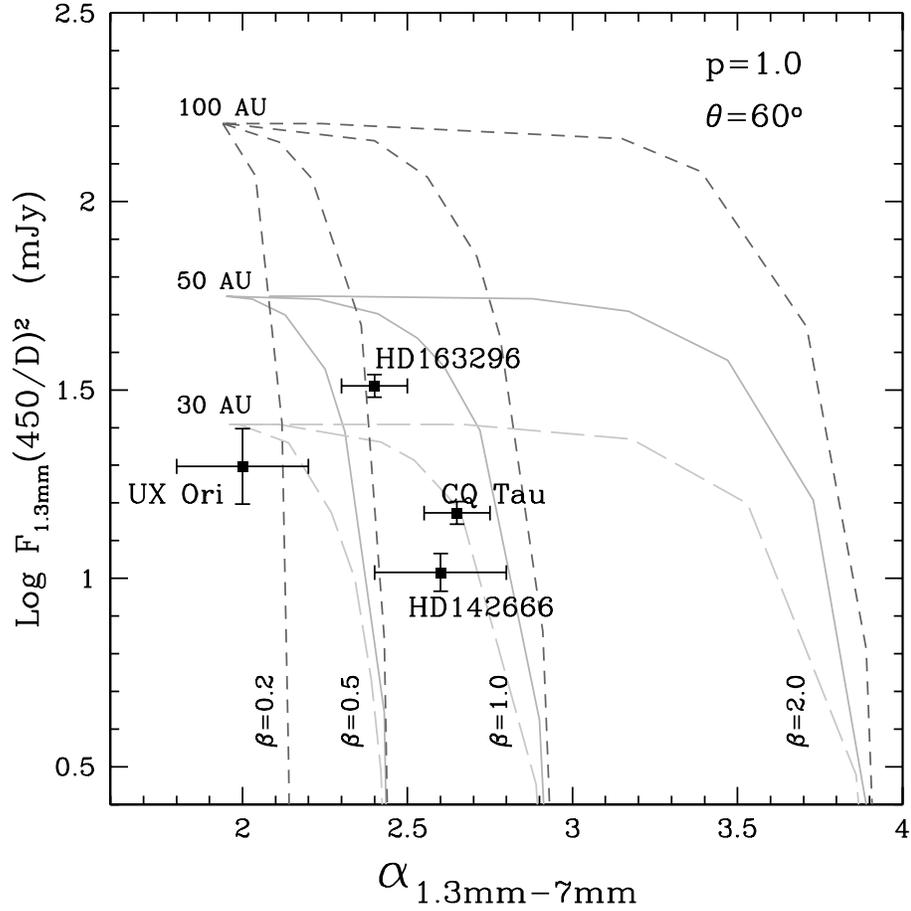}
\caption {Comparison of the observed values  with model predictions
in a plot of the 
1.3mm flux versus the millimeter
spectral index. Each curve is computed for disks having fixed values of the outer radius
$R_D$ and of the dust opacity slope $\beta$, as labelled;  the disk mass decreases
along the curve (from Testi et al.~2001). 
The observed points have been  scaled to take roughly into account
the difference in luminosity of the stars.
}
\end{figure}

It is clear that even for the
extreme case of UX~Ori, which has $\alpha_{1.3mm-7mm}=2 \pm 0.2$,
the observations can be fitted by a very optically thick small disk, or by
a larger, optically thin disk with centimeter size grains (pebbles), 
having $\beta \sim 0$ up to a wavelength of 7mm.
Of the four stars,
CQ~Tau has been marginally resolved both at 1.3mm (Dutrey, personal
communication, quoted in Natta et al.~2001) and at 7mm (Testi et al., in
preparation) and has a rather small disk ($R$ about 50 AU). 
HD~163296 has been marginally
resolved by OVRO (Mannings \& Sargent 1997) with a size of about 100 AU (FWHM)
and an inclination angle of 58 degree. 
Well resolved images at 7mm are necessary to 
discriminate between different models.

So far, the best case for grain growth in a PMS object is that of the
nearby TTS TW~Hya, which has been mapped at 7mm by Wilner et al.~(2000)
and analyzed recently by Calvet et al (2002) using detailed, self-consistent 
disk models. The result is that the fluxes and
7mm surface brightness profile are consistent with an outer disk
where grains have grown to size of $\sim$1 cm.
This, and similar results on other PMS stars that should come in the
near future, will provide much needed observational constraint to 
grain growth and planetesimal formation theories.

\section {Summary}

In this  talk I have presented a summary of the properties of disks 
around pre-main--sequence stars, i.e., when the central objects
have already accreted most of their final mass and have cleared most of the
surrounding material from which they have formed.
This stage of disk evolution is when planets can form. It is, therefore, of
interest  to know their properties and  and how they change during the PMS life
of the stars, if we want to understand how planetary systems form and, in their turn,
evolve.

Pre-MS disks are rich in gas. We know something about
their shape, namely that in many cases they
are ``flared", as expected for gas-rich disks in hydrostatic equilibrium.
Masses are uncertain, and vary by more than two orders of magnitude  for stars
of similar properties. Assuming a gas-to-dust mass ratio of 100, we find
disk masses of 0.003-0.3 M$_\odot$, with a rather constant ratio of disk to stellar
mass of about 0.03, and again a large dispersion. Sizes are a few hundred AU,
with a large spread, due in part to the variety of observational techniques used. 
Temperatures are consistent in most cases with disks being heated by
irradiation from the central star. This is an area of fast progress, as better
disk models ( both of the thermal structure and of the chemistry) as well as
observations of an increasing number of molecular species become available.

Stars of all mass have disks, from objects of $\sim$3-4 M$_\odot$ to brown dwarfs
and planetary-mass objects.  The exception are the more massive stars,
that seem to lack significant disks. This is probably a result of the
fast disk dissipation caused by a combination
of photoevaporation by the stellar radiation of the outer disk and of viscous
evolution of the inner disk.
Less massive stars have typical disk lifetimes of a few million years. However,
there are again large individual variations, and several stars
retain their rather massive PMS disks at ages of 10 Myrs. We do not know
yet what exactly determines the survival of an individual disk, even if we
can guess that being located in a dense cluster or too close to a massive star
is a bad idea.

Finally, I have briefly summarized the state of the research on grain growth
during PMS lifetimes, which is considered the initial step toward
the formation of planetesimals and planets.

\end{document}